\documentclass[fleqn,twoside]{article}
\usepackage{espcrc2}

\readRCS
$Id: espcrc2.tex,v 1.2 2004/02/24 11:22:11 spepping Exp $
\usepackage{graphicx}
\usepackage{epsfig}
\usepackage[figuresright]{rotating}

\newcommand{\lsim}{\raisebox{-0.13cm}{~\shortstack{$<$ \\[-0.07cm] $\sim$}}~}

\title{Associated Higgs production with bottom quarks at hadron colliders}

\author{Michael Kr\"amer\address[EDI]{School of Physics, The University of Edinburgh,
Edinburgh EH9 3JZ, Scotland}%
}

\begin{document}

\begin{abstract}
Higgs-boson production in association with bottom quarks, $\,p\bar p/pp
\to b \bar bH\!+\!X,$ is an important discovery channel for
supersymmetric Higgs particles at the Tevatron and the LHC. We present
higher-order QCD predictions for inclusive cross sections and for the
production of a Higgs boson in association with high-$p_T$ bottom
quarks. We compare calculations performed in a four-flavour scheme
based on the parton processes $gg,q\bar q \to b \bar b H$ with
five-flavour scheme calculations based on bottom-quark scattering.
\vspace{1pc}
\end{abstract}

\maketitle

\section{Introduction}
The Higgs mechanism is a cornerstone of the Standard Model (SM) and
its supersymmetric extensions. The masses of the fundamental
particles, electroweak gauge bosons, leptons, and quarks, are
generated by interactions with Higgs fields. The search for Higgs
bosons is thus one of the most important endeavours in high-energy
physics and is being pursued at the upgraded proton--antiproton
collider Tevatron with a centre-of-mass (CM) energy of $1.96$~TeV,
followed in the near future by the proton--proton collider LHC with
$14$~TeV CM energy.

Various channels can be exploited to search for Higgs bosons at hadron
colliders. Higgs radiation off bottom quarks~\cite{Raitio:1978pt}
\begin{equation}
p\bar p / pp \to b\bar bH\!+\!X
\label{eq:procs_hadron}
\end{equation}
is the dominant Higgs-boson production mechanism in supersymmetric
theories at large $\tan\beta$, where the bottom--Higgs Yukawa coupling
is strongly enhanced.%
\footnote{The parameter $\tan\beta = v_2/v_1$ is the ratio
of the vacuum expectation values of the two Higgs fields generating
the masses of up- and down-type particles in supersymmetric extensions
of the SM. $H = H_{\rm SM}, h^{0}, H^{0}$ may denote the SM Higgs
boson or any of the CP-even neutral Higgs bosons of supersymmetric
theories.}

\section{\boldmath $b\bar{b}H$ production mechanisms}

\begin{figure}[htb]
\vspace*{-9mm}
\epsfig{file=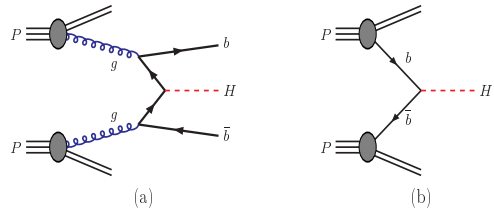,%
        bbllx=75pt,bblly=555pt,bburx=640pt,bbury=755pt,%
        width=8cm,height=3cm,clip=}
\vspace*{-12mm}
\caption[]{\em Generic set of leading order Feynman diagrams for 
$gg \to b\bar bH$(a) and $b\bar b \to H$(b).}
\label{fig:feyn_lo}
\end{figure}

In a four-flavour scheme with no $b$ quarks in the initial state,
the lowest-order QCD processes for associated $b \bar b H$
production are gluon--gluon fusion and quark--antiquark annihilation%
\footnote{The $b\bar b H$ cross section at the Tevatron and the LHC 
is completely dominated by gluon-induced parton processes.}
\begin{equation}
gg \to b\bar bH \quad \mbox{and} \quad q\bar q \to b\bar bH \, , 
\label{eq:procs_parton}
\end{equation}
as shown in Fig.~\ref{fig:feyn_lo}(a).  The inclusive cross section
for $gg \to b\bar bH$ develops potentially large logarithms $\propto
\ln(\mu_F/m_b)$, which arise from the splitting of gluons into nearly 
collinear $b\bar b$ pairs. The large scale $\mu_F\!\!\propto\!\! M_H$
corresponds to the upper limit of the collinear region up to which
factorization is valid. It has been argued that $\mu_F \approx
M_H/4$~\cite{Rainwater:2002hm}. The $\ln(\mu_F/m_b)$ terms can be
summed to all orders in perturbation theory by introducing bottom
parton densities (the five-flavour scheme)~\cite{Barnett:1987jw}. The
five-flavour scheme is based on the approximation that the outgoing
$b$ quarks are at small transverse momentum. In this scheme, the
leading-order (LO) process for the inclusive $b \bar b H$ cross
section is $b\bar b$ fusion, $b\bar b\to H$
(Fig.~\ref{fig:feyn_lo}(b)).
The incoming $b$ partons are given zero transverse momentum at leading
order, and acquire transverse momentum at higher order.  If one
demands that at least one $b$ quark is observed at large transverse
momentum, the leading parton process in the five-flavour scheme is $gb
\to bH$.  A final state with two high-$p_T$ $b$ quarks cannot
be described by $b$-parton densities and has to be calculated through
$gg \to b \bar b H$.

\begin{figure}[htb]
\vspace*{-8mm}
\epsfig{file=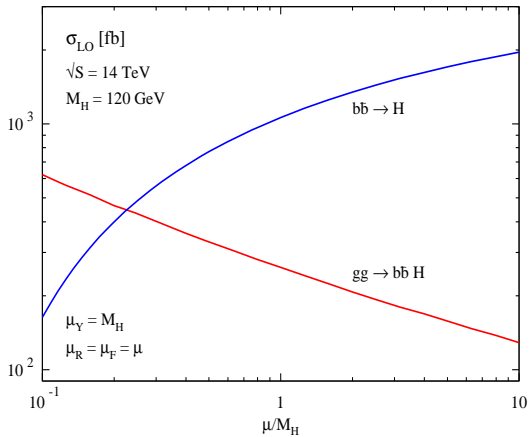,%
        bbllx=30pt,bblly=225pt,bburx=580pt,bbury=635pt,%
        width=7.5cm,height=6cm,clip=}
\vspace*{-13mm}
\caption[]{\em Scale variation of the LO inclusive cross section
 prediction for $pp\to b\bar b H\! +\! X$ at the LHC in the four- and five
 flavour schemes.}
\label{fig:lo_comp}
\vspace*{-8mm}
\end{figure}

To all orders in perturbation theory the four- and five-flavour
schemes are identical, but the way of ordering the perturbative
expansion is different, and the results do not match exactly at finite
order. In Fig.~\ref{fig:lo_comp} we compare the LO predictions for the
total $b \bar b H$ cross section at the LHC in the two schemes. Both
calculations exhibit a strong scale dependence. Fixing the
renormalization and factorization scales to $\mu = M_H$ the
five-flavour scheme prediction exceeds the four-flavour scheme
prediction by more than a factor of five. A similar pattern is found
for the inclusive $b \bar b H$ cross section at the Tevatron.

The four- and five-flavour schemes represent different perturbative
expansions and they are based on different approximations. In the
four-flavour-scheme the $g \to b \bar b$ splitting is calculated
exactly, but large logarithms $\ln(\mu_F/m_b)$, with $\mu_F \propto
M_H$, appear order by order which may spoil the convergence of the
perturbative series. The five-flavour-scheme calculations, on the
other hand, are based on a leading-logarithmic collinear approximation
to the $g \to b \bar b$ splitting which allows to sum the logarithms
$\ln(\mu_F/m_b)$ to all orders by the introduction of evolved $b$
parton densities.

In an attempt to quantify the quality of the approximations in the two
calculational schemes, we have compared the LO cross section for $b
\bar b H$ production in the four- and five-flavour schemes with an
approximate five-flavour scheme calculation where the evolved
$b$-quark parton distribution function is replaced by the calculated
$\alpha_s^1$ contribution to the distribution of heavy quarks in an
on-mass shell gluon:
\begin{eqnarray}
\tilde{b}(x,\mu) & \!  = \! & \frac{\alpha_s(\mu)}{2\pi}
\ln\left( \frac{\mu^2}{m_Q^2}\right)\nonumber \\
&& \times
\int_x^1\frac{d\xi}{\xi} P^{(1)}_{qg}\left(\frac{x}{\xi}\right) 
g(\xi,\mu)\, .
\label{eq:bpdf}
\end{eqnarray}
Here $P^{(1)}_{qg}$ is the usual $\mbox{gluon}\!\! \to\!\! \mbox{quark}$
splitting function $P^{(1)}_{qg}(\xi) = T_F(\xi^2+(1-\xi)^2)$ and
$g(\xi,\mu)$ is the gluon distribution function. Comparing the
four-flavour scheme calculation based on $gg \to b \bar b H$ with the
approximate five-flavour scheme calculation based on $\tilde b
\tilde b \to H$ allows one to quantify the impact of approximating the 
exact $g \to b \bar b$ splitting with the leading-logarithmic
collinear approximation. The difference between the approximate
five-flavour scheme calculation and the five-flavour scheme
calculation with evolved $b$ parton densities provides an estimate of
the effect of summing the $\ln(\mu_F/m_b)$ terms.

\begin{figure}[htb]
\vspace*{-1mm}
\epsfig{file=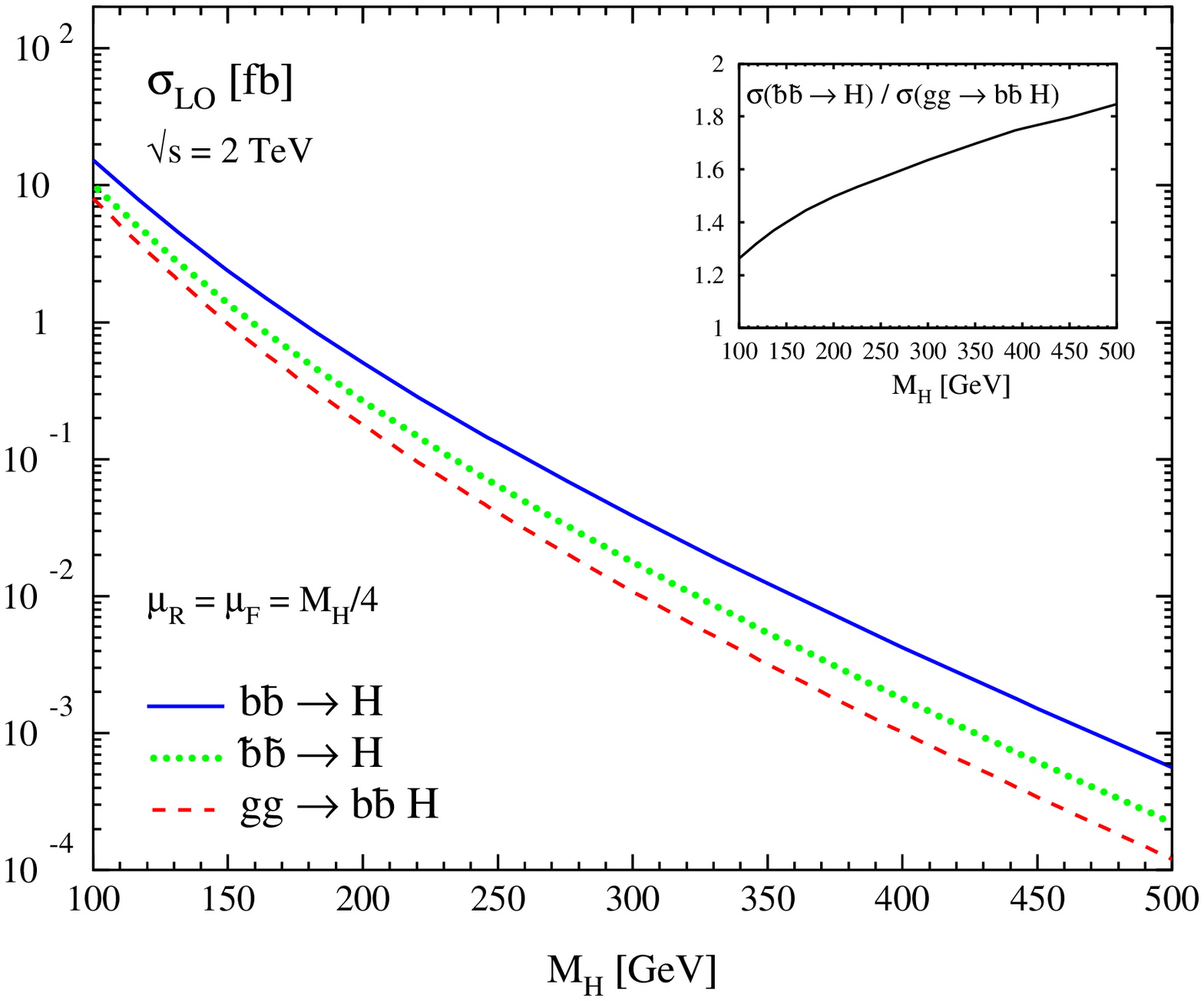,%
        bbllx=30pt,bblly=195pt,bburx=590pt,bbury=655pt,%
        width=7.25cm,height=6cm,clip=}\\
\epsfig{file=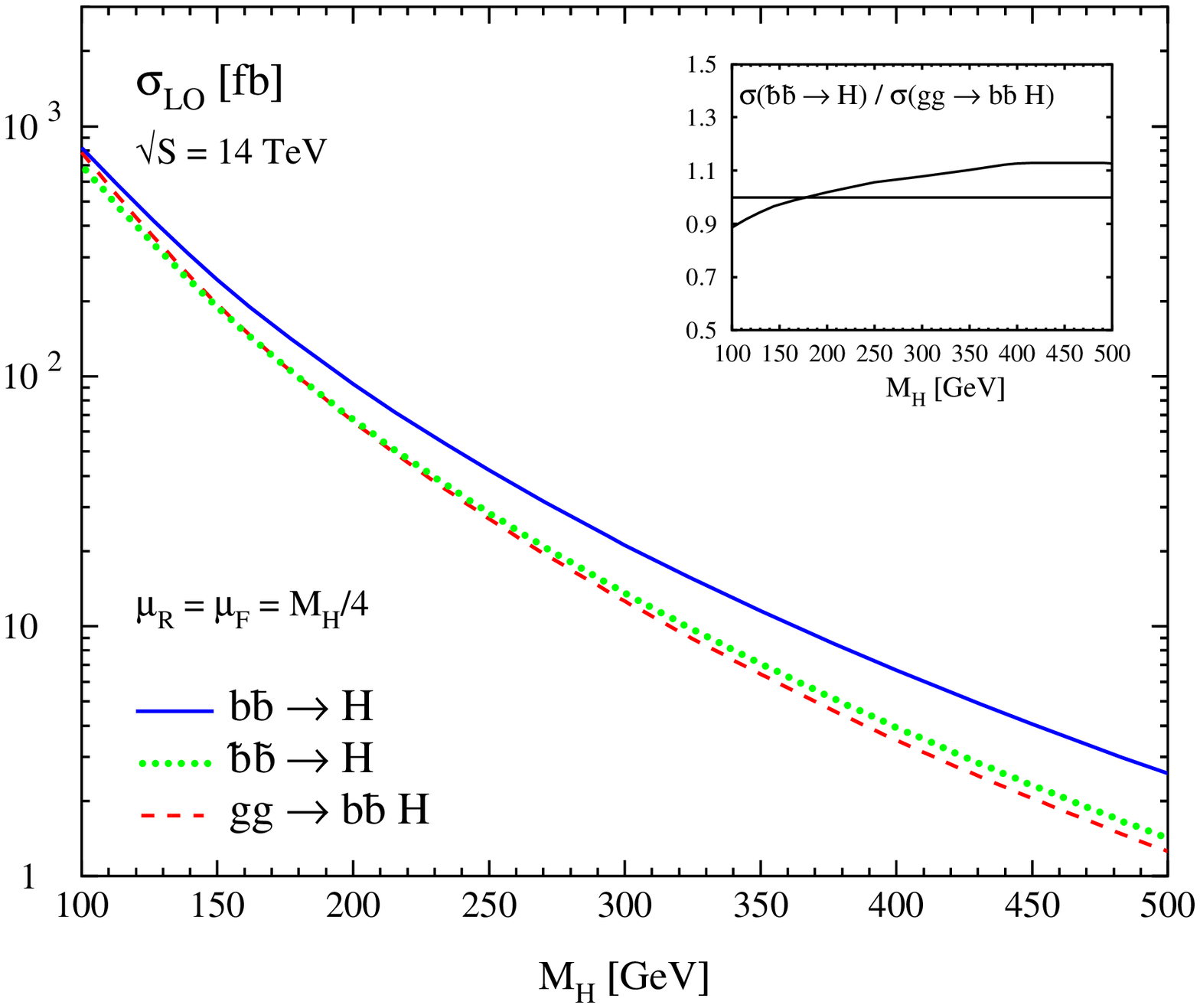,%
        bbllx=30pt,bblly=195pt,bburx=590pt,bbury=655pt,%
        width=7.25cm,height=6cm,clip=}
\vspace*{-11mm}
\caption[]{\em Comparison between the four-, five-, and the approximate
five-flavour scheme calculations at LO for inclusive $p\bar{p}/pp \to
b\bar b H \!+ \!X$ production at the Tevatron and the LHC. The inserts show
the ratio of the approximate five-flavour scheme and the four-flavour
scheme calculations.}
\label{fig:bt_comp}
\vspace*{-8mm}
\end{figure}

The comparison between the four-, five-, and the approximate
five-flavour scheme calculations is presented in
Fig.~\ref{fig:bt_comp}. We have set all scales equal to $\mu = M_H/4$,
which is an appropriate factorization scale choice in the five-flavour
scheme~\cite{Rainwater:2002hm,Harlander:2003ai}. The results shown in
Fig.~\ref{fig:bt_comp} imply that the leading logarithmic
approximation to the $g \to b \bar b$ splitting is not very accurate
for $b \bar b H$ production at the Tevatron. The LO cross-section
prediction based on the $\tilde b \tilde b \to H$ process exceeds the
exact calculation through $gg \to b \bar b H$ by up to a factor 1.5
for $M_H\lsim 200$~GeV. At the same time, the effect of summing the
$\ln(\mu_F/m_b)$ terms is significant, increasing the cross section
prediction by up to a factor 1.7 in the same Higgs mass
range. Fortunately, the comparison between the different calculational
schemes is more favourable for $b \bar b H$ production at the LHC. The
leading logarithmic approximation to the $g \to b \bar b$ splitting
function reproduces the exact calculation within approximately 10\%,
while the effect of summing the $\ln(\mu_F/m_b)$ terms is less than
about 40\% for $M_H\lsim 200$~GeV.

As we shall demonstrate in Section~\ref{sec:comp}, the scheme dependences
are reduced when higher-order QCD corrections are included. However,
the LO analysis presented in Fig.~\ref{fig:bt_comp} suggests that the
approximations involved in both the four- and five-flavour scheme
calculations are not very accurate for $b \bar b H$ production at the
Tevatron. At the LHC, on the other hand, both schemes should yield
reliable and compatible results.

\vspace*{-0.5mm}
\section{Higher-order QCD corrections}
\label{sec:nlo}

The inclusion of higher-order QCD corrections is crucial to reduce the
scale and scheme dependence of the LO cross-section predictions. The
five-flavour scheme $b \bar b \to H$ process has been calculated to
next-to-leading order (NLO)~\cite{Dicus:1998hs} and
next-to-next-to-leading order (NNLO)~\cite{Harlander:2003ai}
accuracy. The NNLO corrections strongly reduce the renormalization and
factorization scale dependences, see also Ref.~\cite{harlander:ll2004}.

\begin{figure}[htb]
\vspace*{-1mm}
\epsfig{file=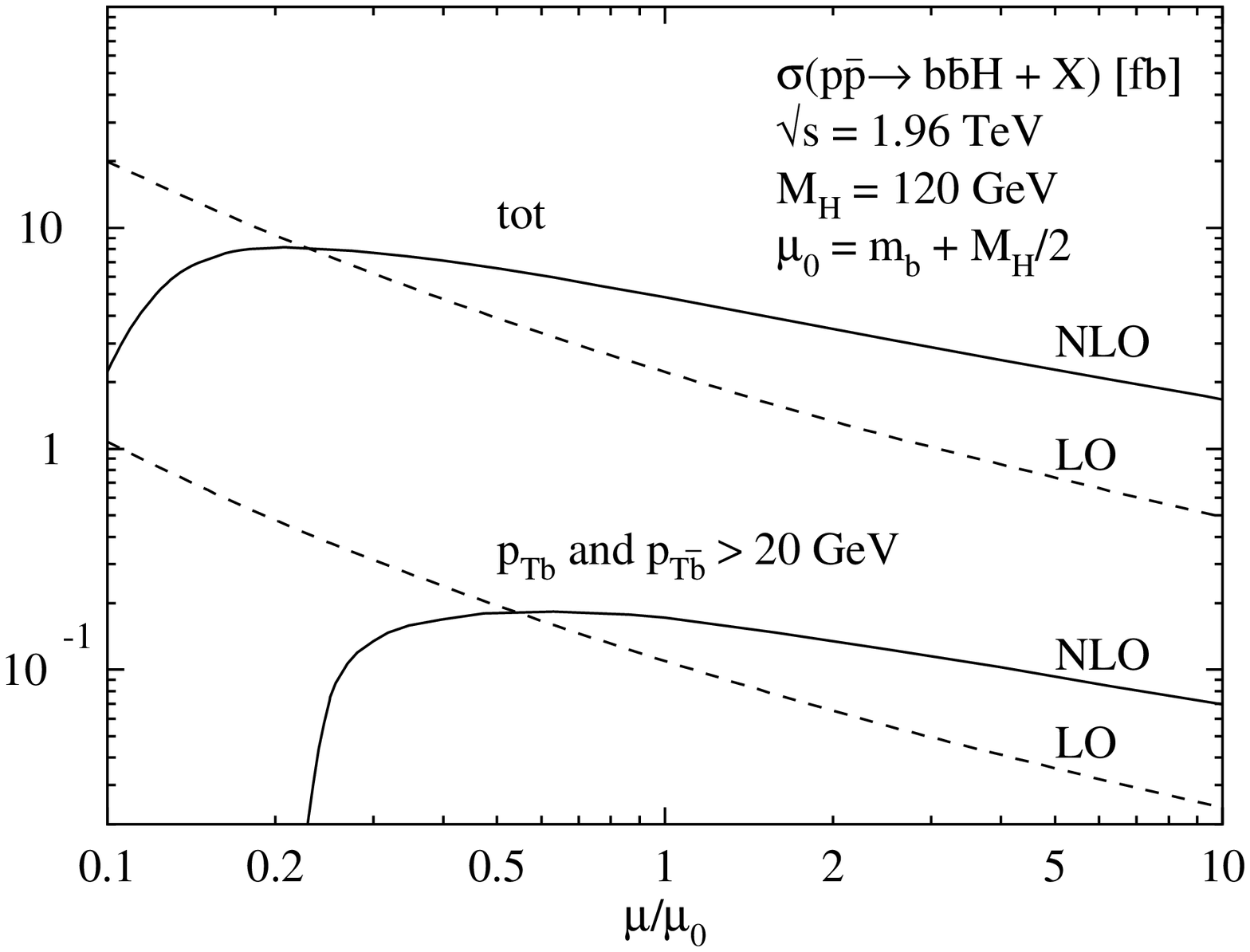,%
        bbllx=20pt,bblly=220pt,bburx=580pt,bbury=650pt,%
        width=7.5cm,height=6cm,clip=}\\
\epsfig{file=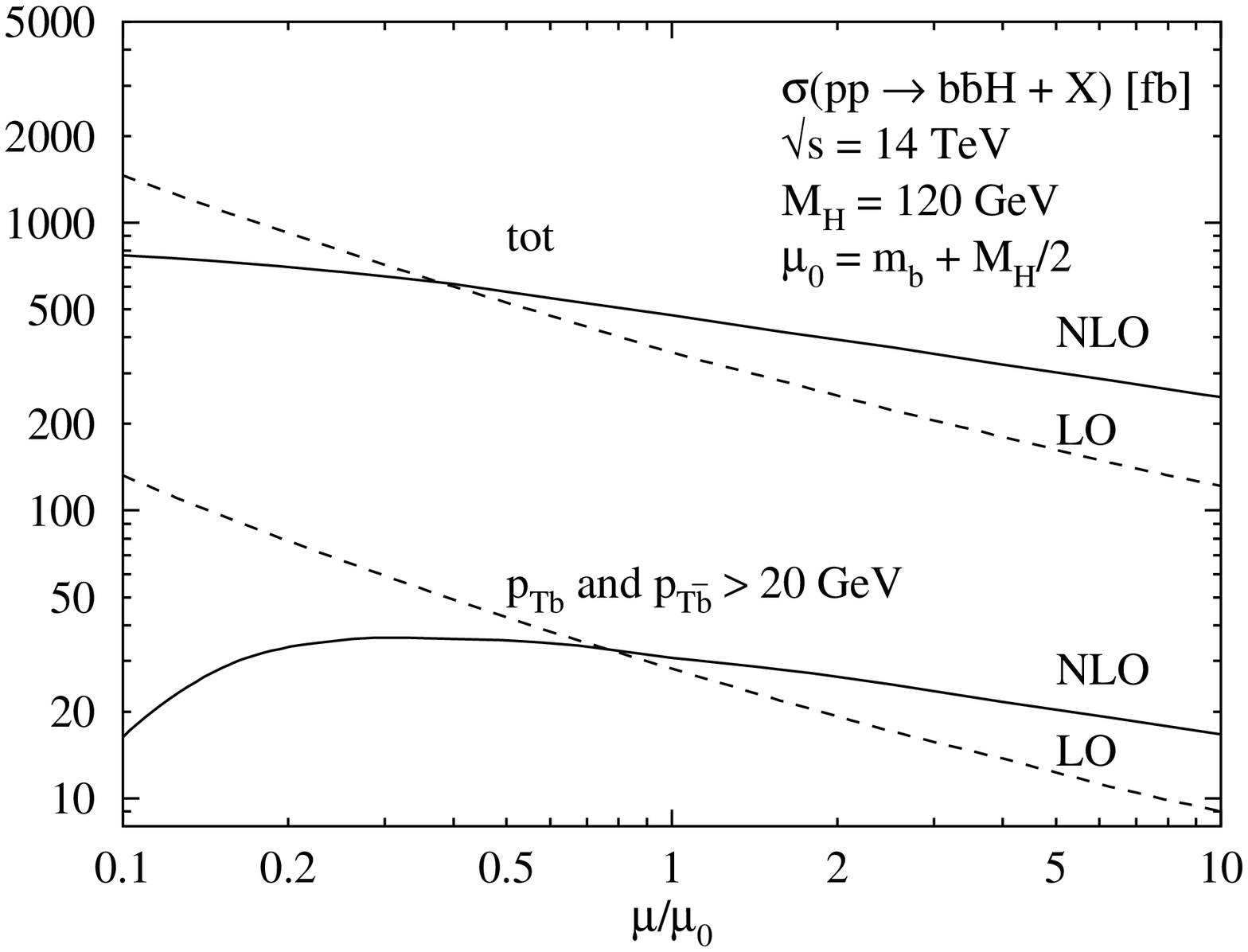,%
        bbllx=20pt,bblly=220pt,bburx=580pt,bbury=650pt,%
        width=7.5cm,height=6cm,clip=}
\vspace*{-12mm}
\caption[]{\em Scale variation of the NLO cross section prediction for
 $p\bar{p}/pp\to b\bar b H\! +\! X$ at the Tevatron and the LHC in the
 four-flavour scheme. From Ref.~\cite{Dittmaier:2003ej}.}
\label{fig:scale}
\vspace*{-9mm}
\end{figure}

The calculation of the NLO QCD corrections to the four-flavour-scheme
processes $gg, q\bar q \to Q \bar Q H$, where $Q$ denotes a generic
heavy quark, has been described in Ref.~\cite{Beenakker:2001rj}.  NLO
results for the total $b \bar b H$ cross
section~\cite{Dittmaier:2003ej} and for the production of a Higgs
boson in association with high-$p_T$ $b$
quarks~\cite{Dittmaier:2003ej,Dawson:2003kb} have been presented in
the literature. Figure~\ref{fig:scale} shows the LO and NLO scale
dependence for the total cross section and for the cross section with
two high-$p_T$ $b$ quarks~\cite{Dittmaier:2003ej}. The reduction of
the scale dependence at NLO is particularly significant for the
exclusive cross section where both $b$ quarks are required to be
produced with large transverse momentum.

\vspace*{-0.5mm}
\section{Comparison of 4- and 5-flavour schemes}
\label{sec:comp}

\begin{figure}[htb]
\vspace*{-1mm}
\epsfig{file=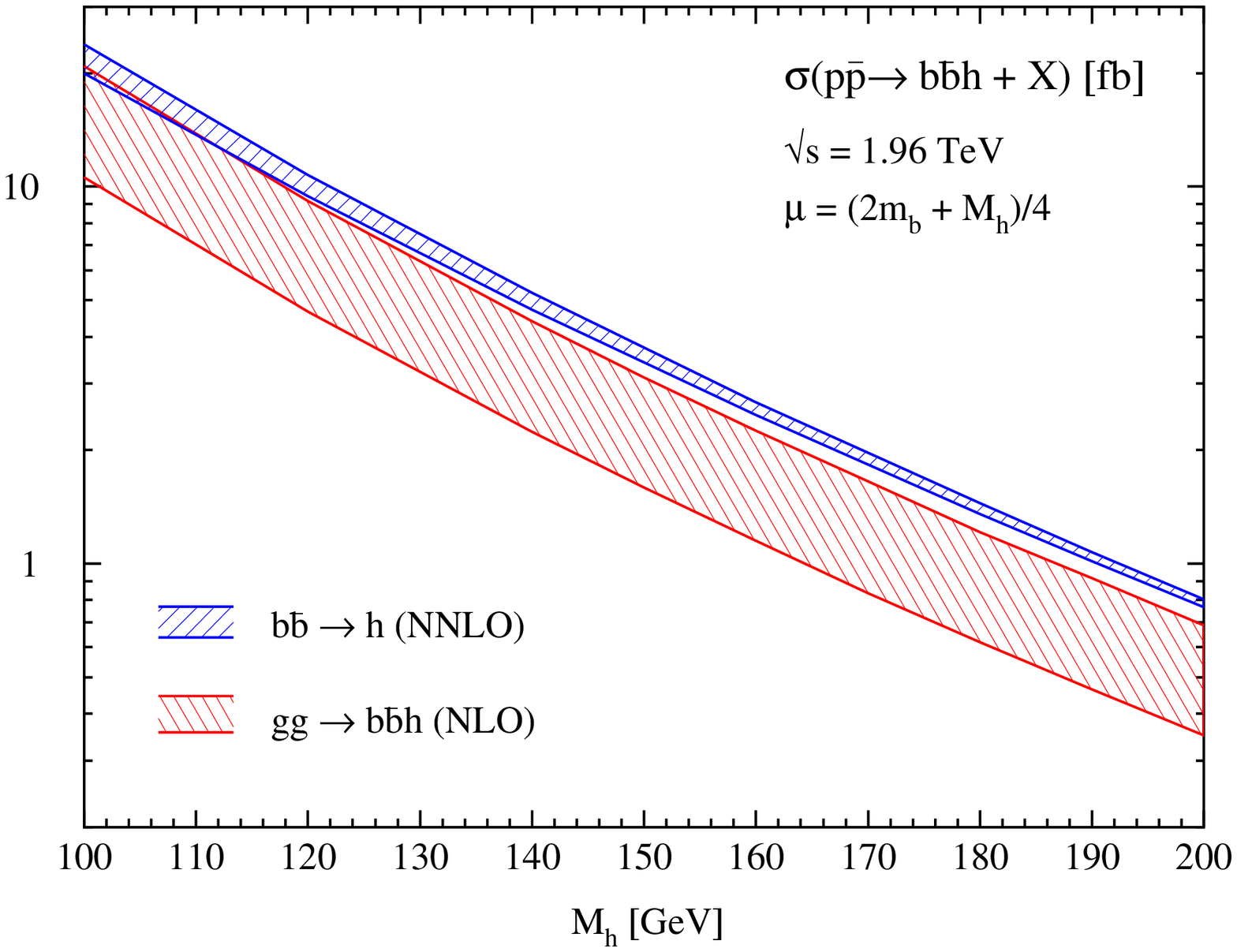,%
        bbllx=20pt,bblly=220pt,bburx=580pt,bbury=650pt,%
        width=7.5cm,height=6cm,clip=}\\
\epsfig{file=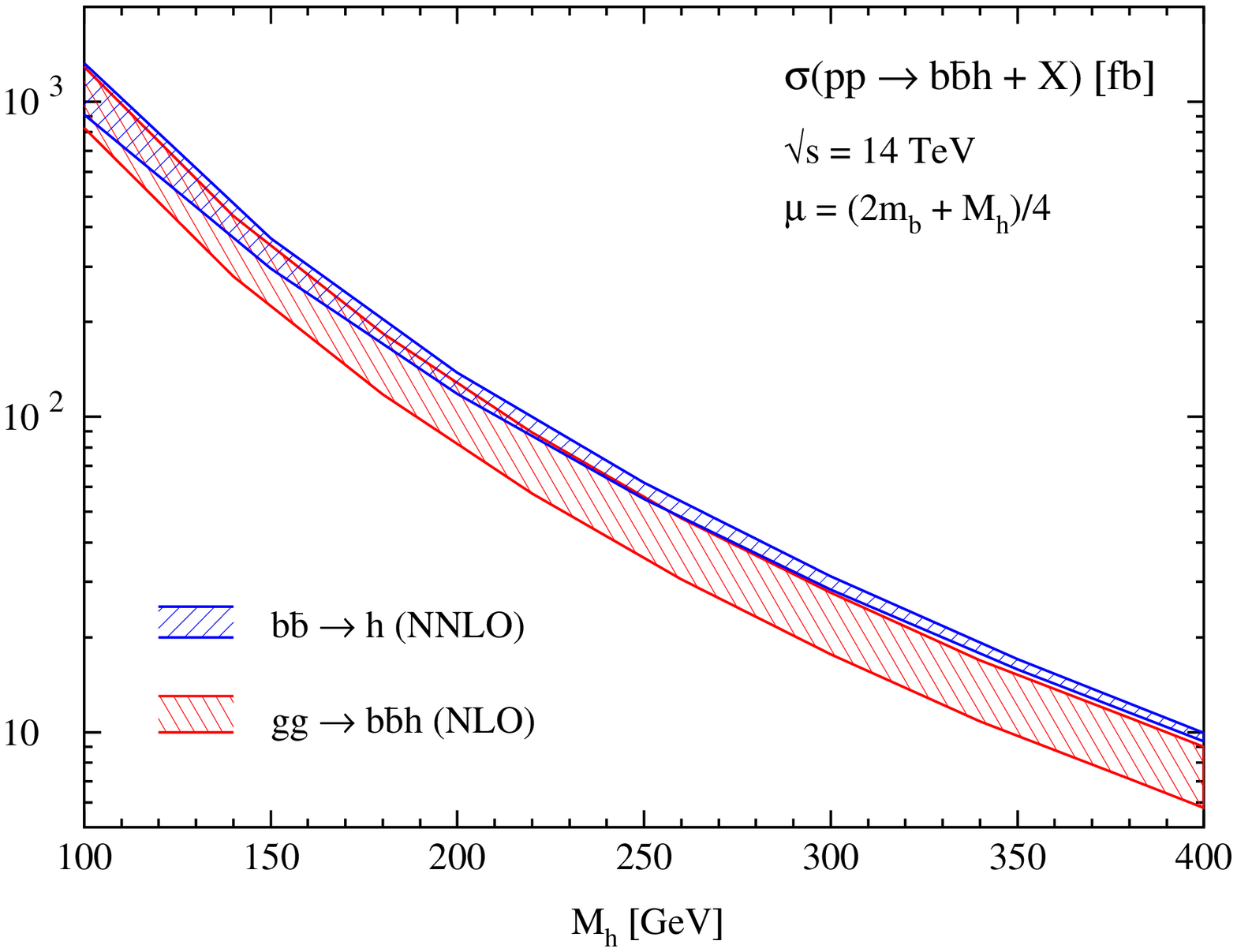,%
        bbllx=20pt,bblly=220pt,bburx=580pt,bbury=650pt,%
        width=7.5cm,height=6cm,clip=}
\vspace*{-12mm}
\caption[]{\em Inclusive cross section for $p\bar p/pp \to b \bar b H
\!+\! X$ at the Tevatron and the LHC in the four-flavour scheme at
NLO~\cite{Dittmaier:2003ej} and the five-flavour scheme at
NNLO~\cite{Harlander:2003ai}. From Ref.~\cite{Campbell:2004pu}.}
\label{fig:xs_comp}
\vspace*{-8mm}
\end{figure}

Despite the sizable scale uncertainty at NLO, the four-flavour-scheme
calculation yields a reliable prediction for the inclusive cross
section. This is demonstrated in Fig.~\ref{fig:xs_comp} where the NLO
four-flavour scheme calculation is compared with the NNLO calculation
of $b \bar b \to H$ (five-flavour scheme). The two calculations are
compatible within their respective scale uncertainties for small Higgs
masses, while for large Higgs masses the five-flavour scheme tends to
yield larger cross sections. As suggested by the LO analysis,
Fig.~\ref{fig:bt_comp}, the comparison between the two schemes is more
favourable for $b\bar b H$ production at the LHC.%
\footnote{Note that Higgs radiation off closed top-quark loops has not
been included in the NNLO calculation of $b \bar b \to H$. These
contributions are negligible in supersymmetric extensions of the SM at
large $\tan\beta$, but they reduce the SM prediction in the
four-flavour scheme by $\approx 10\%$.}

\begin{figure}[htb]
\vspace*{-1mm}
\epsfig{file=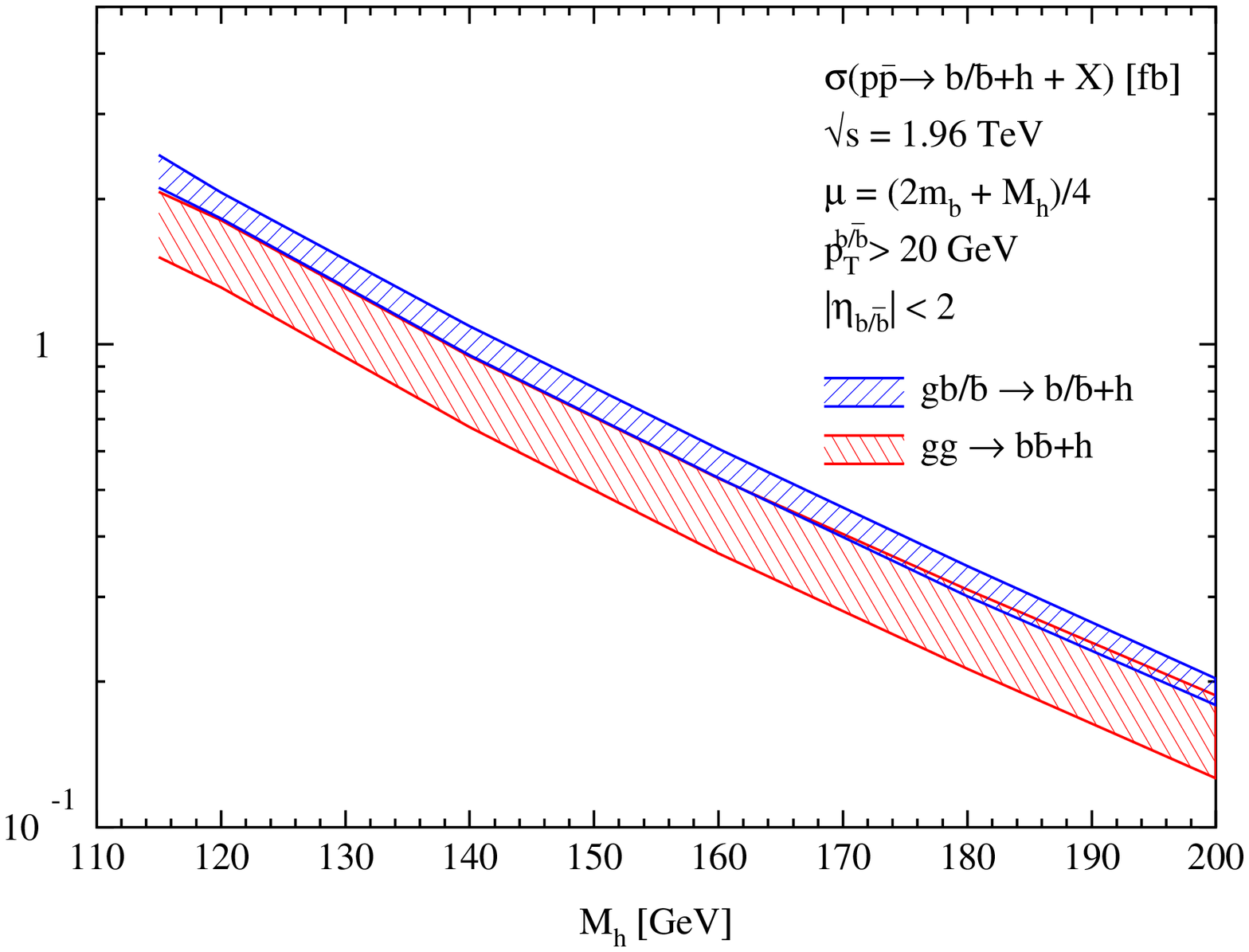,%
        bbllx=20pt,bblly=220pt,bburx=580pt,bbury=650pt,%
        width=7.25cm,height=6cm,clip=}\\
\epsfig{file=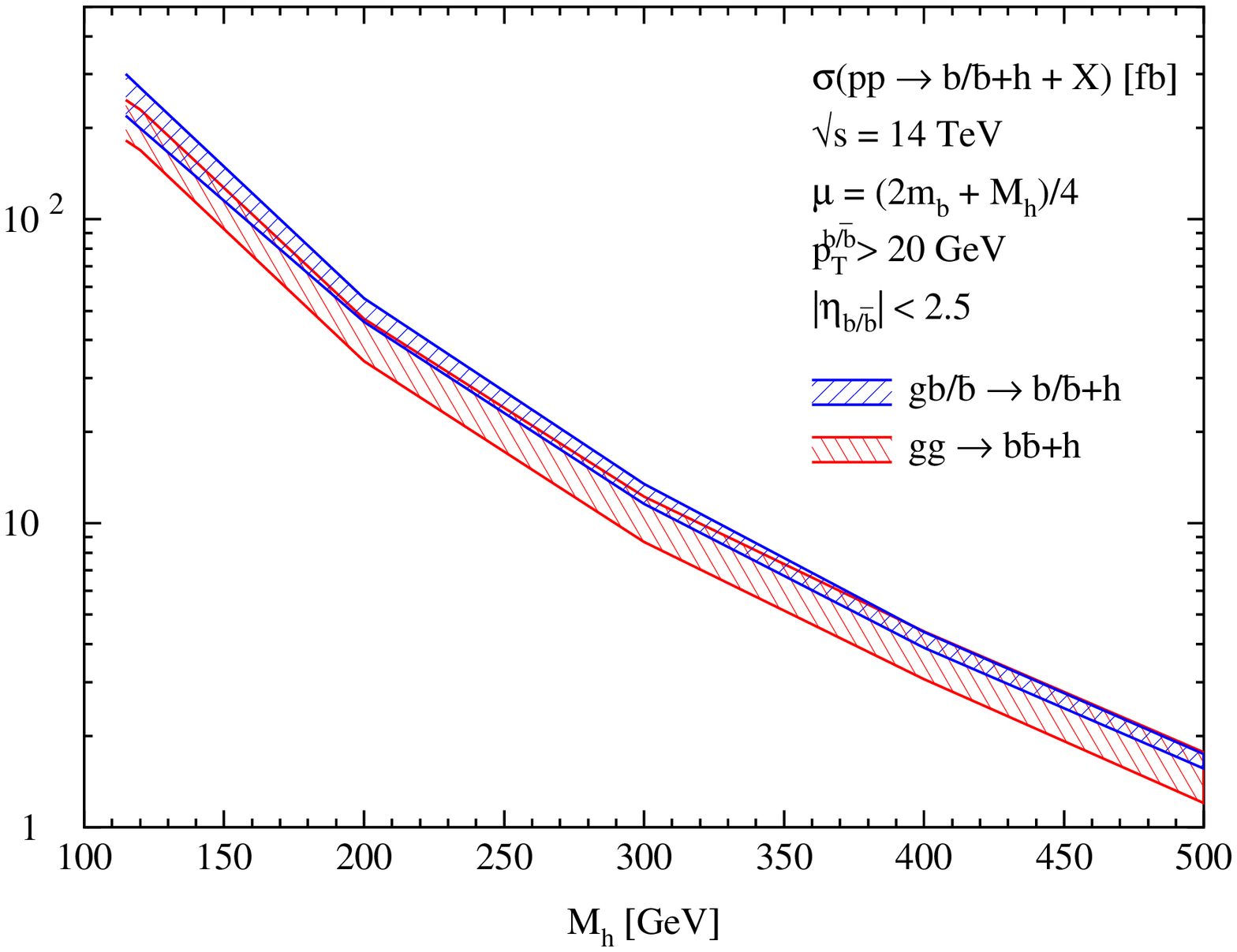,%
        bbllx=20pt,bblly=220pt,bburx=580pt,bbury=650pt,%
        width=7.25cm,height=6cm,clip=}
\vspace*{-10mm}
\caption[]{\em NLO cross section for $p\bar p/pp \to b \bar b H\! +\! X$
with one high-$p_T$ $b$ quark at the Tevatron and the LHC in the
four-flavour~\cite{Dittmaier:2003ej} and
five-flavour~\cite{Campbell:2002zm} schemes. From
Ref.~\cite{Campbell:2004pu}.}
\label{fig:1tag_comp}
\vspace*{-7mm}
\end{figure}

Requiring a high-$p_T$ $b$ quark in the final state reduces the signal
cross section with respect to the inclusive cross section, but the $b$
quark can be used to suppress the background and to identify the
Higgs-boson production mechanism. Figure~\ref{fig:1tag_comp} shows the
NLO cross section predictions for the production of a Higgs boson plus
a single $b$ quark. Results are compared between the four-flavour
scheme based on the parton processes $gg,q\bar q \to b \bar b H$ with
the momentum of one of the $b$ quarks integrated
over~\cite{Dittmaier:2003ej}, and the five-flavour scheme based on the
process $g b \to b H$~\cite{Campbell:2002zm}. As for the inclusive
cross section, the two approaches agree within their scale
uncertainty, but the five-flavour scheme tends to yield larger cross
sections.

\vspace*{-4mm}
\section{Conclusions}
Associated $b\bar b H$ production at the Tevatron and the LHC is an
important discovery channel for Higgs bosons at large values of
$\tan\beta$ in supersymmetric extensions of the SM, where the bottom
Yukawa coupling is strongly enhanced. Results in the four- and
five-flavour schemes have been compared, including higher-order QCD
corrections. The two calculational schemes represent different
perturbative expansions of the same physical process, and therefore
should agree at sufficiently high order.  It is satisfying that the
NLO (and NNLO) calculations for $b\bar b H$ production are compatible
within their uncertainties.  This is a significant advance over
several years ago, when comparisons of $b\bar b\to h$ at NLO and
$gg\to b\bar bh$ at LO showed large discrepancies.
 
\vspace*{2mm}

\noindent
{\bf Acknowledgements}\\ I am grateful to S.~Dittmaier, M.~Spira and
the authors of Ref.~\cite{Campbell:2004pu} for their collaboration. I
would like to thank the organisers of Loops and Legs 2004 for their
kind invitation.

\end{document}